\documentclass[aps,prb,reprint,superscriptaddress,showpacs=false]{revtex4-2}
\usepackage{graphicx}
\usepackage{amsmath}
\usepackage{siunitx}
\usepackage{booktabs}
\usepackage{xcolor}
\usepackage[colorlinks=true,allcolors=blue]{hyperref}

\newcommand{\EbAAhollow}{-1.92}  
\newcommand{\MAAhollow}{0.00}  
\newcommand{\gapAAhollow}{3.32}  
\newcommand{\dEAAbridge}{636}  
\newcommand{\MAAbridge}{4.38}  
\newcommand{\gapAAbridge}{4.11}  
\newcommand{\dEAAtop}{665}  
\newcommand{\MAAtop}{4.48}  
\newcommand{\EbABmid}{-1.25}  
\newcommand{\MABmid}{4.32}  
\newcommand{\gapABmid}{3.99}  
\newcommand{\dEABtopb}{103}  
\newcommand{\dEABtop}{158}  
\newcommand{\dEmonohollow}{9}  
\newcommand{\dEmonobridge}{59}  
\newcommand{\Ebmonooffhollow}{-0.60}  
\newcommand{\Mmonooffhollow}{5.75}  
\newcommand{\gapHostAA}{3.57}  
\newcommand{\dEABmidShiftMeV}{0.27}  
\newcommand{\dEABmidtwoMeV}{-0.11}  
\newcommand{\dEABmidtwoShiftMeV}{0.36}  
\newcommand{\dEABmidMzeroEV}{1.22}  
\newcommand{\dEABhollowTopMeV}{-0.1}  
\newcommand{\ABmidSeedSpreadMeV}{0.1}  
\newcommand{\AAhollowSeedSpreadMeV}{0.2}  
\newcommand{\dEmonoFMtwoEV}{1.33}  
\newcommand{\dEmonoFMfourEV}{0.58}  
\newcommand{\monoMinTwinMeV}{-0.1}  
\newcommand{\dEmonoHollowTwoMeV}{0.07}  
\newcommand{\dEAAbridgeUhalfMeV}{284}  
\newcommand{\dEAAbridgeUoneMeV}{-74}  
\newcommand{\dEAAhollowPinFreeFiveMeV}{14}  
\newcommand{\dEAAnoDthreeMeV}{-224}  
\newcommand{\dEstackHostFiveMeV}{342}  
\newcommand{\dEstackCrFiveFreeMeV}{96}  
\newcommand{\dEstackHostThreeMeV}{98}  
\newcommand{\dEbStackThreeMeV}{666}  
\newcommand{\dEbStackFourFreeMeV}{263}  
\newcommand{\dEbStackFourPinMeV}{531}  
\newcommand{\dEbStackFiveFreeMeV}{246}  
\newcommand{\dEbStackUthreeMeV}{107}  
\newcommand{\AAbridgeSeedSpreadMeV}{0.2}  
\newcommand{\GateMaxdEMeV}{9.7\times10^{-6}}  
\newcommand{\GPAWversion}{25.7.0}  
\newcommand{\ASEversion}{3.29.0}  
\newcommand{\EaAAscanThree}{636}  
\newcommand{\fAAcrossLoThree}{0.20}  
\newcommand{\fAAcrossHiThree}{0.35}  
\newcommand{\EaAAscanFour}{372}  
\newcommand{\fAApeakFour}{0.25}  
\newcommand{\fAAcrossLoFour}{0.20}  
\newcommand{\fAAcrossHiFour}{0.25}  
\newcommand{\EaAAscanFive}{260}  
\newcommand{\fAApeakFive}{0.20}  
\newcommand{\MAApeakFive}{4.6}  
\newcommand{\fAAcrossLoFive}{0.15}  
\newcommand{\fAAcrossHiFive}{0.20}  
\newcommand{\EaAAscanFourFree}{236}  
\newcommand{\fAApeakFourFree}{0.20}  
\newcommand{\fAAcrossLoFourFree}{0.15}  
\newcommand{\fAAcrossHiFourFree}{0.20}  
\newcommand{\EaAAscanFiveFree}{185}  
\newcommand{\fAApeakFiveFree}{0.20}  
\newcommand{\fAAcrossLoFiveFree}{0.15}  
\newcommand{\fAAcrossHiFiveFree}{0.20}  
\newcommand{\EaAAinterpMaxDevMeV}{11}  
\newcommand{\AApeakWhereThree}{the bridge (magnetic plateau)}  
\newcommand{\AApeakWhereFour}{the spin-crossing region}  
\newcommand{\AApeakWhereFive}{the spin-crossing region}  
\newcommand{\dEAAbridgeFourPin}{314}  
\newcommand{\dEAAbridgeFivePin}{245}  
\newcommand{\dEAAbridgeFourFree}{223}  
\newcommand{\dEAAbridgeFiveFree}{183}  
\newcommand{\dEAAfarFourMeV}{135}  
\newcommand{\dEAAfarFiveMeV}{75}  
\newcommand{\KinT}{300}  %
\newcommand{\EaABNEB}{15}  
\newcommand{\logkABlo}{12}  
\newcommand{\logkABhi}{13}  
\newcommand{\EamonoNEB}{62}  
\newcommand{\logkmonolo}{11}  
\newcommand{\logkmonohi}{12}  
\newcommand{\EaABNEBseven}{13}  
\newcommand{\monoNEBsevenSteps}{69}  
\newcommand{\monoNEBsevenFmaxLo}{0.03}  
\newcommand{\monoNEBsevenFmaxHi}{0.19}  
\newcommand{\monoNEBsevenHiMeV}{72}  
\newcommand{\monoNEBsevenBelowMeV}{6}  
\newcommand{\EamonoNEBfromLowestMeV}{68}  
\newcommand{\dEmonoBridgeHollowFourMeV}{-42}  
\newcommand{\dEmonoTopBridgeFourMeV}{13}  
\newcommand{\dEmonoSegAFourMeV}{5}  
\newcommand{\dEmonoSegBFourMeV}{12}  
\newcommand{\dEmonoHopMaxFourMeV}{13}  
\newcommand{\dEmonoBridgeHollowFiveMeV}{-39}  
\newcommand{\dEmonoTopBridgeFiveMeV}{10}  
\newcommand{\dEmonoSegAFiveMeV}{2}  
\newcommand{\dEmonoSegBFiveMeV}{9}  
\newcommand{\dEmonoHopMaxFiveMeV}{10}  
\newcommand{\dEmonoHollowminFiveMeV}{10}  
\newcommand{\dxyHollowminFive}{0.6}  
\newcommand{\monoBridgeMinMaxMeV}{0.3}  
\newcommand{\dEmonoBridgeHollowThreeMeV}{50}  
\newcommand{\MmonoBridgeFive}{5.6}  
\newcommand{\dEmonoHollowBridgeFiveMeV}{39}  
\newcommand{\dEmonoTopBridgeThreeMeV}{2.5}  
\newcommand{\logkmonoFivelo}{12}  
\newcommand{\logkmonoFivehi}{13}  
\newcommand{\logkAAThreelo}{1}  
\newcommand{\logkAAThreehi}{2}  
\newcommand{\dlogkABAAThree}{10}  
\newcommand{\logkAAFivelo}{8}  
\newcommand{\logkAAFivehi}{9}  
\newcommand{\dlogkABAAFive}{4}  
\newcommand{\logkAAFiveFreelo}{9}  
\newcommand{\logkAAFiveFreehi}{10}  
\newcommand{\dlogkABAAFiveFree}{3}  

\begin{document}

\title{Confinement, registry and spin in chromium migration between graphene layers}

\author{Mingguang Chen}
\email[Corresponding author: ]{deepgroundingai@gmail.com}
\affiliation{DeepGrounding}
\author{Senfu Zhang}
\email[Corresponding author: ]{zhangsenfu@lzu.edu.cn}
\affiliation{Key Laboratory of Magnetism and Magnetic Functional Materials (Lanzhou University), Ministry of Education, Lanzhou University, Lanzhou 730000, China}

\date{\today}

\begin{abstract}
We compare the migration of a single Cr atom on monolayer graphene, between AB-stacked layers and between AA-stacked layers, with one first-principles protocol (PBE+D3(BJ), one cell family, registry anchors and spin-state search). We find three main results.
First, on the surface and in the AB gallery Cr stays magnetic and moves almost freely. On the surface the C--C bridge is the minimum in $4\times4$ and $5\times5$ cells, and the highest point we sample on a hop lies \dEmonoHopMaxFiveMeV~meV above it ($5\times5$). In the AB gallery Cr sits on the inversion centres of the bilayer, which form a kagome net, and hops between them over \EaABNEB~meV with a moment of about \MABmid~$\mu_\mathrm{B}$.
Second, in the AA gallery the minimum is a nonmagnetic $\eta^6$--$\eta^6$ sandwich that pulls the layers together. Leaving it requires opening the gallery and switching Cr to a moment of 4--5~$\mu_\mathrm{B}$, so we map the path with separately relaxed spin branches. The barrier estimate is \EaAAscanThree~meV in a $3\times3$ cell but \EaAAscanFive{} and \EaAAscanFiveFree~meV in $5\times5$ cells with the far-field spacing pinned and free, and it is not yet converged. At 300~K, with equal prefactors and if the spin change is not rate-limiting, Cr hops between AA layers \dlogkABAAFiveFree--\dlogkABAAFive{} orders of magnitude more slowly than between AB layers, using the $5\times5$ AA estimates and the $3\times3$ AB band. These AA numbers are specific to PBE+D3: an on-site $U$ of 1.0~eV on Cr $d$ already makes the magnetic states the AA ground state.
Third, Cr binds \dEbStackFiveFreeMeV--\dEbStackThreeMeV~meV more strongly in AA than in AB registry, so the fast AB hop requires that the layers keep their Bernal registry around Cr.
\end{abstract}

\maketitle

\section{Introduction}

Chromium forms bis-hexahapto ($\eta^6$-$\eta^6$) complexes that bind two graphitic $\pi$ surfaces, and this chemistry has been used to interconnect graphene and carbon-nanotube networks electrically~\cite{sarkar2011chemsci,chen2016chemmater,chen2017ucr}. During vapour deposition of Cr onto graphene--nanotube--graphene devices, conductance changes appear region by region rather than simultaneously~\cite{chen2017ucr}. That observation raises an atomistic question: how mobile is an isolated Cr atom on a graphene surface compared with Cr confined between two graphene layers, and what does confinement do to its spin?

Several first-principles studies address parts of this question, each with a different protocol.
\begin{itemize}
\item Hardcastle \emph{et al.} computed Cr adatom migration on the exterior surfaces of mono-, bi- and trilayer graphene and found Cr to be highly mobile~\cite{hardcastle2013prb}. The accompanying thesis lists hollow$\to$bridge$\to$hollow barriers of 0.021--0.022~eV for Cr~\cite{hardcastle2014thesis}.
\item Han, Kang and Dai reported a 0.74~eV climbing-image barrier for one Cr atom between AA-stacked layers in a $6\times6$ cell (PBE with D2 dispersion)~\cite{han2018rscadv}.
\item Liao \emph{et al.} listed 0.14~eV for single-Cr diffusion in AB-stacked bilayer graphene~\cite{liao2016pccp}.
\end{itemize}
The electronic structure of ordered Cr-intercalated bilayers has been studied separately~\cite{li2024nanoscale,chen2026tmbg}. Because these barriers were obtained with different codes, cells, dispersion corrections, constraints and spin treatments, the fivefold difference between the AA and AB gallery values cannot be attributed to stacking alone.

Here we compute surface, AA-gallery and AB-gallery migration of a single Cr atom with one protocol. The functional, dispersion correction, cutoff, cell family, registry constraint and spin-state search are the same for all three, so that differences can be attributed to confinement and registry. We follow the Cr moment along each path. In the AA gallery the spin state changes during the hop, so we map that path with separately relaxed spin branches rather than with a single collinear band. We also test the numerical and methodological sensitivity of the results, including the cell size. We do not claim the first calculation of Cr migration in bilayer graphene. Our contribution is a consistent comparison, the spin evolution along the paths, the dependence of the AA barrier on cell size and far-field treatment, and an explicit account of what remains uncertain.

\section{Methods}
\label{sec:methods}

\subsection{Electronic structure}
Calculations use GPAW~\cite{gpaw,gpaw2024} \GPAWversion{} in plane-wave mode with ASE~\cite{ase} \ASEversion{}.
\begin{itemize}
\item \emph{Functional and dispersion.} We use the PBE functional~\cite{pbe} with the D3 dispersion correction~\cite{d3} and Becke--Johnson damping~\cite{d3bj}.
\item \emph{Basis and setups.} The plane-wave cutoff is 450~eV, and the default GPAW PAW setups are used. All calculations are spin-polarized and collinear, without spin--orbit coupling.
\item \emph{Smearing.} Occupations use Fermi--Dirac smearing of 0.05~eV for bilayers and 0.10~eV for the monolayer adatom. With Cr on the $4\times4$ monolayer, self-consistency at 0.05~eV had not converged after 38 iterations (MixerDif) or 106 iterations (MixerSum) when the tests were stopped; at 0.10~eV it converged in 105 iterations. The pristine monolayer reference uses the same 0.10~eV, and the effect of the smearing on the surface barrier is checked at fixed geometry (Sec.~\ref{sec:sens}).
\item \emph{Brillouin zone.} Sampling is $\Gamma$-centred, with $4\times4\times1$ points for the $3\times3$ supercell and $3\times3\times1$ for the $4\times4$ and $5\times5$ supercells. Point-group symmetry is disabled.
\item \emph{Convergence.} SCF thresholds are $10^{-5}$~eV per valence electron for the energy, $10^{-4}$ for the density, $10^{-8}$~eV$^2$ per electron for the eigenstates and 0.005~eV/\AA{} for the forces. Geometries are relaxed with BFGS to a maximum force of 0.02~eV/\AA.
\item \emph{Hubbard correction.} The PBE+U checks add $U=0.5$, 1.0, 1.5 or 3~eV on Cr $d$ through the GPAW setup option.
\end{itemize}

\subsection{Models and constraints}
Supercells are $n\times n$ graphene cells ($n=3,4,5$) with lattice constant 2.467~\AA{} and 16~\AA{} of vacuum along $z$. Bilayers are either AA-stacked (eclipsed) or AB-stacked (Bernal). One Cr atom per cell is placed between the layers (``gallery'') or above a monolayer (``surface'').

\begin{figure*}
\includegraphics[width=\textwidth]{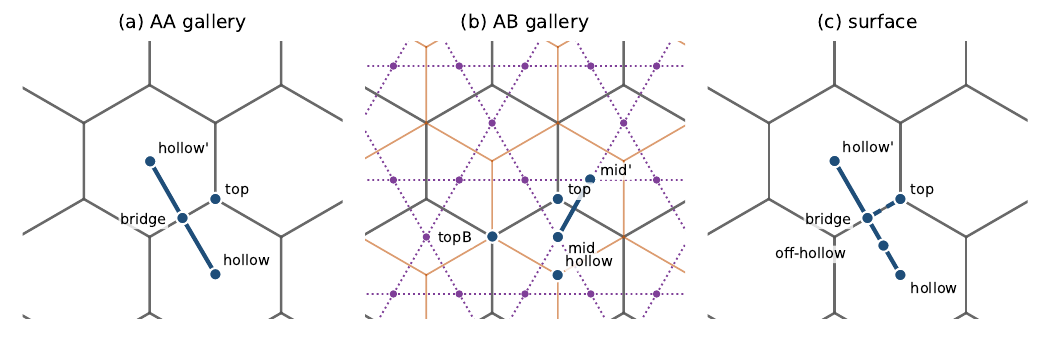}
\caption{\label{fig:sites}Top views of the $3\times3$ cells. Grey: lower graphene layer (in AA the upper layer is directly above it); orange: upper layer of AB. Blue points are relaxed Cr positions taken from the run records, and blue lines the CI-NEB paths. (a) AA gallery, hollow$\to$bridge$\to$hollow$^\prime$. (b) AB gallery. The purple dotted net joins the inversion centres of the bilayer that do not lie on an atom pair. It is computed from the relaxed pristine host and forms a kagome net. The hop is mid$\to$mid$^\prime$ between neighbouring centres. (c) Surface. Solid: the $3\times3$ CI-NEB path. Dashed: the bridge$\to$top segment sampled in the larger cells (Sec.~\ref{sec:sens}).}
\end{figure*}

The lateral sites are labelled as follows:
\begin{itemize}
\item hollow: hexagon centre of the lower layer;
\item bridge: C--C bond midpoint;
\item top: above a lower-layer carbon;
\item topB: in AB stacking, above a lower-layer carbon that carries an upper-layer carbon.
\end{itemize}
In AB stacking, the midpoint between a top and a hollow site is an inversion centre of the bilayer. We call it ``mid'' (Fig.~\ref{fig:sites}), and an equivalent inversion centre half a lattice vector away ``mid$^\prime$''. We checked numerically that inverting the pristine AB structure about mid maps every carbon onto a carbon to within $10^{-14}$~\AA.

\emph{Registry anchors.} One far-field carbon in each layer has its in-plane coordinates fixed. This prevents rigid sliding and translation without biasing the local relaxation. All other coordinates are free unless stated, so AA results are at fixed AA registry.

\emph{Site energies.} Some site energies are obtained by relaxing with the Cr in-plane coordinates held on the site; the height and all host coordinates are free. We use this constraint for the AA scan points, the $5\times5$ bridges, the free $4\times4$ AA bridge, and the larger-cell surface top sites and points between bridge and top. The other sites were relaxed without it. Without the constraint Cr can slide away from a site whose symmetry the cell does not fully preserve, and some of our early larger-cell runs did so. The script that builds the tables therefore checks every unconstrained relaxation and rejects any run in which Cr moved more than 0.1~\AA{} in plane. Minimum searches are exempt from this check. A site energy difference, such as bridge minus hollow, is not a migration barrier. Barriers in this paper come only from the CI-NEB bands, the AA scan envelope and the sampled surface hop, and each is labelled as such.

\emph{Far-field spacing.} In the $4\times4$ and $5\times5$ cells we use two far-field treatments.
\begin{itemize}
\item ``Free'': all host heights relax.
\item ``Pinned'': carbon heights beyond a lateral radius of 4.5~\AA{} ($4\times4$) or 5.0~\AA{} ($5\times5$) from the cell centre are held at the relaxed pristine interlayer spacing. Every Cr site is at least 3.8~\AA{} from a pinned carbon.
\end{itemize}
In a periodic cell with one Cr, the free treatment lets the whole gallery open, whereas the pinned treatment confines the opening to the neighbourhood of Cr. Neither is guaranteed to bound the dilute limit, so we report both and treat their spread as an uncertainty.

\subsection{Spin states}
Sites were relaxed from several initial Cr moments between 0 and 6~$\mu_\mathrm{B}$; these are seeds, not assumptions. For the sites discussed below, the seeds converge to a single state within \AAhollowSeedSpreadMeV~meV (AA hollow), \AAbridgeSeedSpreadMeV~meV (AA bridge) and \ABmidSeedSpreadMeV~meV (AB mid). States that a free relaxation does not reach were computed with the total moment fixed ($M=0$, 2 or 4~$\mu_\mathrm{B}$).

Binding energies are $E_b = E(\text{host+Cr}) - E(\text{host}) - E(\text{Cr})$. The isolated Cr atom is computed in its high-spin ($^7S$, 6~$\mu_\mathrm{B}$) state in a 14~\AA{} box. Runs that fail self-consistency are reported as failures and not used.

\subsection{Migration paths}
\emph{Surface and AB gallery.} Paths on the surface and in the AB gallery keep their spin state (Table~\ref{tab:neb}). For these we use climbing-image nudged elastic band (CI-NEB) calculations~\cite{henkelman2000cineb,henkelman2000tangent} with three moving images, converged to 0.05~eV/\AA. As a check on the discretisation, we recomputed both bands with seven moving images and a 0.02~eV/\AA{} threshold, started from the converged three-image bands with a midpoint inserted between neighbouring images. The AB band converged; the surface band did not (Sec.~\ref{sec:paths}).
\begin{itemize}
\item \emph{AB.} The endpoints are the relaxed final geometries of the two mid-site stability tests (Sec.~\ref{sec:sites}), and the initial band uses IDPP interpolation~\cite{smidstrup2014idpp}. The interlayer gap is nearly the same along this hop.
\item \emph{Surface.} The initial band is piecewise linear through the separately relaxed bridge.
\end{itemize}
We avoided IDPP wherever the gap must change along the path. IDPP keeps the gap of the endpoints, so in a test band for the AA gallery it squeezed Cr at the midpoint and raised that image by several eV.

Each image has its own spin seed. We check that the band endpoints reproduce the separately relaxed minima; $\delta_\mathrm{end}$ in Table~\ref{tab:neb} is the largest deviation.

\emph{AA gallery.} The AA minimum is nonmagnetic, whereas the bridge is magnetic (Sec.~\ref{sec:sites}). A collinear CI-NEB band started from per-image seeds did not represent the change: its climbing image stayed on the magnetic branch and drifted toward the hollow. We therefore scan the hollow$\to$bridge segment, which lies on a mirror line of the AA bilayer. The Cr in-plane position is held at hollow $+f\,(\text{bridge}-\text{hollow})$, every other coordinate is relaxed, and each spin branch is relaxed separately:
\begin{itemize}
\item $M=0$: total moment fixed at 0, starting from the relaxed hollow;
\item magnetic: free moment, seeded with 4 or 5~$\mu_\mathrm{B}$, starting from the relaxed bridge or from a relaxed magnetic scan point;
\item in the $4\times4$ cell also $M=2$: total moment fixed at 2~$\mu_\mathrm{B}$.
\end{itemize}
The barrier estimate $E_\mathrm{env}$ is the highest point of the lower envelope of these branches at the sampled $f$. It is a sampled value, not a bound. The crossing between branches is only bracketed by the sampled points, and between two samples the envelope can lie above both. As a check on the sampling, we also report the crossing of straight lines through the two lowest branches at the bracketing points. $E_\mathrm{env}$ also does not locate the minimum-energy crossing point between the spin surfaces, and paths off the mirror line may be lower.

\subsection{Rates}
Hop rates at $T=\KinT$~K are estimated as $k=\nu\exp(-E_a/k_BT)$ with $\nu$ between $10^{12}$ and $10^{13}$~s$^{-1}$. We did not compute the prefactor~\cite{vineyard1957}; the range is a typical one.

\subsection{Hosts and reproducibility}
Runs were distributed over three Linux hosts: two aarch64 and one x86\_64, all with the same GPAW version and byte-identical PAW setup files.
\begin{itemize}
\item \emph{Gate.} Before any result from a host was used, the host ran four fixed-geometry statics that were compared with the reference host. The largest energy difference was $\GateMaxdEMeV$~meV.
\item \emph{Run records.} Every task is defined by a JSON file. Each result file records the code and package versions, the host, the number of MPI ranks, and SHA-256 hashes of the task and source files.
\item \emph{Tables and figures.} These are generated by one script from the result files, and each quoted number lists its source runs (Sec.~\ref{sec:data}).
\end{itemize}

\section{Results}

\subsection{Sites, binding and spin}
\label{sec:sites}
\begin{table*}
\caption{\label{tab:sites}Relaxed Cr sites in the $3\times3$ cells. $E_b$: binding energy; $\Delta E$: energy relative
to the lowest site of the same host; $M$: total cell moment; $M_\mathrm{Cr}$: moment in the Cr PAW sphere; CN: carbon
neighbours within 2.4~\AA; $d_\mathrm{min}$: shortest Cr--C distance; gap: mean interlayer spacing.}
\begin{ruledtabular}
\begin{tabular}{lrrrrrrr}
Site & $E_b$ (eV) & $\Delta E$ (meV) & $M$ ($\mu_\mathrm{B}$) & $M_\mathrm{Cr}$ ($\mu_\mathrm{B}$) & CN & $d_\mathrm{min}$ (\AA) & gap (\AA) \\
\hline
AA hollow & -1.92 & 0 & 0.00 & 0.00 & 12 & 2.20 & 3.32 \\
AA bridge & -1.28 & 636 & 4.38 & 3.58 & 4 & 2.21 & 4.11 \\
AA top & -1.25 & 665 & 4.48 & 3.56 & 2 & 2.13 & 4.10 \\
AB mid & -1.25 & 0 & 4.32 & 3.47 & 6 & 2.16 & 3.99 \\
AB topB & -1.15 & 103 & 4.43 & 3.53 & 2 & 2.12 & 4.07 \\
AB top & -1.09 & 158 & 4.25 & 3.37 & 7 & 2.12 & 3.82 \\
AB hollow & -1.10 & 158 & 4.25 & 3.37 & 7 & 2.12 & 3.83 \\
mono hollow & -0.59 & 9 & 5.75 & 4.07 & 0 & 2.83 & -- \\
mono bridge & -0.54 & 59 & 5.74 & 4.02 & 2 & 2.38 & -- \\
mono off-hollow & -0.60 & 0 & 5.75 & 4.08 & 0 & 2.65 & -- \\
\end{tabular}
\end{ruledtabular}
\end{table*}

Table~\ref{tab:sites} lists the relaxed sites in the $3\times3$ cells. The three environments bind Cr very differently.

\emph{AA gallery.} The hollow is the minimum, with $E_b=\EbAAhollow$~eV. Cr sits between two eclipsed hexagons and is nonmagnetic ($M=\MAAhollow~\mu_\mathrm{B}$) for every seed. The two layers close in to $\gapAAhollow$~\AA{} around it, compared with $\gapHostAA$~\AA{} in the pristine AA bilayer. This is the bis-hexahapto sandwich of the molecular chemistry.

The bridge and top sites are \dEAAbridge~meV and \dEAAtop~meV higher. They are magnetic ($M=\MAAbridge$ and $\MAAtop~\mu_\mathrm{B}$), and the gap opens to $\gapAAbridge$~\AA.

\emph{AB gallery.} No site offers a hexagon above a hexagon. The minimum is the inversion centre (mid): $E_b=\EbABmid$~eV, $M=\MABmid~\mu_\mathrm{B}$, gap $\gapABmid$~\AA. The topB site lies \dEABtopb~meV higher. The top and hollow sites lie \dEABtop~meV higher and differ from each other by \dEABhollowTopMeV~meV, as the inversion symmetry requires.

We tested that mid is a minimum rather than a saddle:
\begin{itemize}
\item displacing Cr by about 0.07~\AA{} and relaxing without constraints returns it to the site, with an energy change of \dEABmidShiftMeV~meV;
\item mid$^\prime$, with Cr held in plane, lies \dEABmidtwoMeV~meV from mid, and a displaced start there also relaxes back (\dEABmidtwoShiftMeV~meV);
\item a nonmagnetic state at mid, obtained with the total moment fixed at zero, is \dEABmidMzeroEV~eV higher.
\end{itemize}

\emph{Surface.} Cr is weakly bound ($E_b=\Ebmonooffhollow$~eV) and keeps almost its free-atom moment ($M=\Mmonooffhollow~\mu_\mathrm{B}$). The minimum is not the hexagon centre. The CI-NEB band (Sec.~\ref{sec:paths}) showed images displaced from the hollow toward a bridge lying below the hollow, and a free relaxation from such an image gives the minimum ``off-hollow'' in Table~\ref{tab:sites}. A second relaxation, started 0.05~\AA{} off the hollow$\to$bridge mirror line, ends \monoMinTwinMeV~meV from the first. The centred hollow lies \dEmonohollow~meV above it and the bridge \dEmonobridge~meV above it. The two inequivalent hollows of the cell agree to \dEmonoHollowTwoMeV~meV. Closer to the sheet (Cr 1.7~\AA{} above the hollow), states with the total moment fixed at 4 and 2~$\mu_\mathrm{B}$ lie \dEmonoFMfourEV~eV and \dEmonoFMtwoEV~eV above the minimum. Relaxations from those geometries with a free moment failed self-consistency and are not used. Within this search we find no low-spin surface state close to the high-spin minimum. These statements refer to the $3\times3$ cell. In the larger cells the ordering of the surface sites changes (Sec.~\ref{sec:sens}).

\subsection{Surface and AB-gallery paths}
\label{sec:paths}
\begin{table*}
\caption{\label{tab:neb}Climbing-image NEB results ($3\times3$; three moving images converged to 0.05~eV/\AA, and seven
moving images converged to 0.02~eV/\AA{} where stated). $E_a$: highest image minus the
lowest of the band images and any separately relaxed minimum on the path;
$\delta_\mathrm{end}$: largest deviation of a band endpoint from the separately relaxed endpoint; $M$: total moment of
each image along the band. The surface value is a $3\times3$ result and is not converged with cell size
(Sec.~\ref{sec:sens}).}
\begin{ruledtabular}
\begin{tabular}{llrrrl}
Host & Hop & $E_a$ (meV) & $\delta_\mathrm{end}$ (meV) & steps & $M$ along band ($\mu_\mathrm{B}$) \\
\hline
AB & mid$\to$mid$^\prime$ & 15 & 0.0 & 21 & 4.3, 4.4, 4.4, 4.3, 4.3 \\
mono & hollow$\to$hollow & 62 & 0.2 & 22 & 5.8, 5.8, 5.7, 5.8, 5.8 \\
AB & mid$\to$mid$^\prime$, 7 images & 13 & 0.0 & 19 & 4.3, 4.3, 4.3, 4.4, 4.4, 4.4, 4.3, 4.3, 4.3 \\
\end{tabular}
\end{ruledtabular}
\end{table*}

\begin{figure}
\includegraphics[width=\columnwidth]{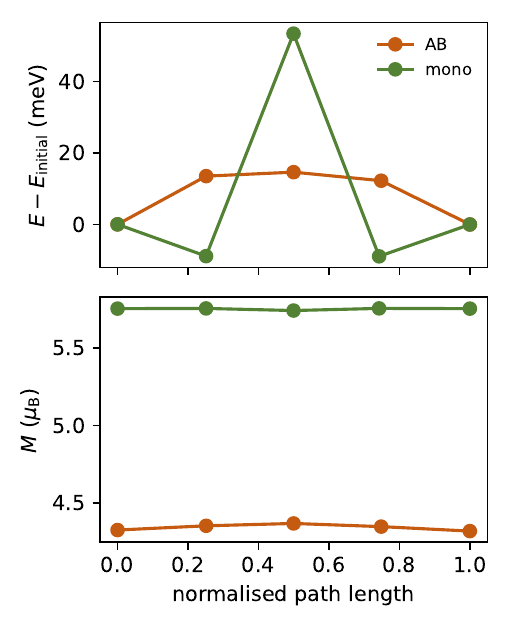}
\caption{\label{fig:neb}Energy (top) and total moment (bottom) along the CI-NEB bands for the surface and AB-gallery hops ($3\times3$). Energies are relative to the first image. On the surface the images next to the endpoints lie below them because the centred hollow is not the minimum (Sec.~\ref{sec:sites}). The surface band is a $3\times3$ result; the surface landscape changes in larger cells (Sec.~\ref{sec:sens}).}
\end{figure}

Both paths keep their spin state, so a single collinear band describes them (Table~\ref{tab:neb}, Fig.~\ref{fig:neb}).

\emph{Surface.} The band was started between the two hollows of the cell, with the highest image at the bridge. The barrier, measured from the relaxed off-hollow minimum, is \EamonoNEB~meV, consistent with the relaxed bridge site (\dEmonobridge~meV). The potential near the hexagon centre is very flat. The seven-image band for this hop did not converge: after \monoNEBsevenSteps{} steps its largest force still varied between \monoNEBsevenFmaxLo{} and \monoNEBsevenFmaxHi~eV/\AA, and its highest image (\monoNEBsevenHiMeV~meV) was not at the bridge, so we do not use it as a barrier. Its lowest image lies \monoNEBsevenBelowMeV~meV below the off-hollow minimum; measured from that point, the three-image barrier would be \EamonoNEBfromLowestMeV~meV. The $3\times3$ surface barrier is therefore uncertain by several meV. This $3\times3$ value is not converged with cell size. In the $4\times4$ and $5\times5$ cells the bridge is the minimum, and the highest point sampled on a bridge$\to$top$\to$bridge hop lies only \dEmonoHopMaxFourMeV{} and \dEmonoHopMaxFiveMeV~meV above it (Sec.~\ref{sec:sens}). The larger-cell value is of the same magnitude as the 0.02~eV of Hardcastle \emph{et al.}~\cite{hardcastle2014thesis}, but the landscapes differ. In their calculation the hollow is the minimum and the bridge the saddle, whereas in our larger cells the bridge is the minimum and the hollow lies \dEmonoHollowBridgeFiveMeV~meV above it and is not a local minimum ($5\times5$). The two calculations also differ in code and dispersion correction (CASTEP with Tkatchenko--Scheffler~\cite{hardcastle2013prb}). All of these barriers are at most a few $k_BT$ at room temperature, so Cr is effectively free to move on the surface.

\emph{AB gallery.} Cr moves from mid to the neighbouring inversion centre mid$^\prime$ (a hop of $a/2=1.23$~\AA) over \EaABNEB~meV. With seven moving images and a 0.02~eV/\AA{} threshold the barrier is \EaABNEBseven~meV (Table~\ref{tab:neb}). It keeps a moment of about $\MABmid~\mu_\mathrm{B}$ throughout, and the gap stays near $\gapABmid$~\AA. The inversion centres at the three $M$ points of the primitive cell are related by the threefold axes of Bernal stacking and form a kagome net, so this hop alone connects the minima in two dimensions.

A hop that passes over the topB site would instead cost at least \dEABtopb~meV. Liao \emph{et al.} list 0.14~eV for Cr in AB bilayer graphene, but the main text does not state the cell or path~\cite{liao2016pccp}, so we cannot tell which of these routes their value corresponds to.

\subsection{AA-gallery path: opening the gallery and switching spin}
\label{sec:aa}
\begin{figure*}
\includegraphics[width=\textwidth]{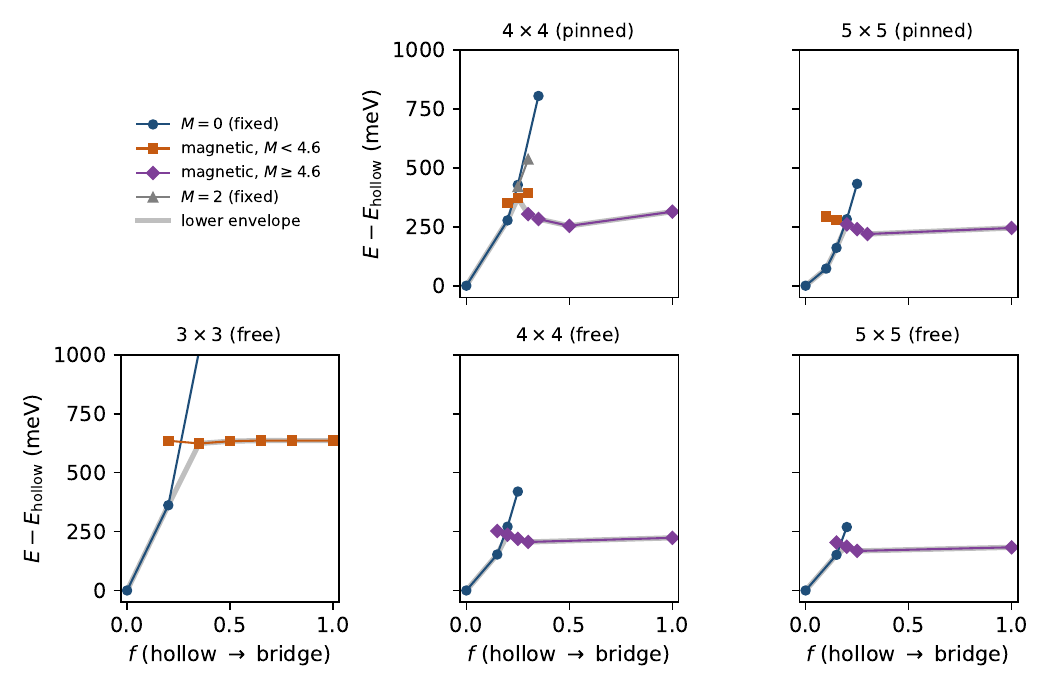}
\caption{\label{fig:aa}AA gallery: energy along the hollow ($f=0$) to bridge ($f=1$) segment, with Cr held in plane and each spin branch relaxed separately. Top row: $4\times4$ and $5\times5$ cells with the far-field spacing pinned. Bottom row: $3\times3$, $4\times4$ and $5\times5$ cells with a free far field. Magnetic points are split by total moment. Lines join sampled points and are not interpolations; the grey line is the lower envelope. The $M=0$ branch of the $3\times3$ cell continues above the plotted range.}
\end{figure*}
\begin{table*}
\caption{\label{tab:aa}AA gallery: dependence of the hollow$\to$hollow barrier estimate on cell size and far-field treatment
(meV). ``bridge'': relaxed bridge minus relaxed hollow, a site energy difference, not a barrier, in the far field given (``pinned'': carbons beyond the pinning radius held at the pristine
spacing; ``free'': all host heights relaxed). $E_\mathrm{env}$: maximum over the sampled points of the lower envelope of the
separately relaxed spin branches along the hollow$\to$bridge segment, at fraction $f$, with the branch and total moment
($\mu_\mathrm{B}$) there; $f_\times$: sampled bracket of the change of the lowest branch from $M=0$ to magnetic.
Cr is held in plane on the hollow$\to$bridge line. $E_\mathrm{env}$ is a sampled value on that line, not a bound to the
minimum-energy path. Rows with the same far field isolate the cell size, and rows with the same cell isolate the far field.}
\begin{ruledtabular}
\begin{tabular}{llrrrlll}
Cell & far field & bridge & $E_\mathrm{env}$ & $f$ & branch, $M$ & $f_\times$ & points \\
\hline
$3\times3$ & free & 636 & 636 & 0.65 & mag., 4.4 & 0.20--0.35 & 10 \\
$4\times4$ & pinned & 314 & 372 & 0.25 & mag., 4.0 & 0.20--0.25 & 13 \\
$4\times4$ & free & 223 & 236 & 0.20 & mag., 4.8 & 0.15--0.20 & 7 \\
$5\times5$ & pinned & 245 & 260 & 0.20 & mag., 4.6 & 0.15--0.20 & 9 \\
$5\times5$ & free & 183 & 185 & 0.20 & mag., 4.7 & 0.15--0.20 & 5 \\
\end{tabular}
\end{ruledtabular}
\end{table*}

\emph{3$\times$3 cell.} Moving Cr out of the AA hollow costs energy on both branches (Fig.~\ref{fig:aa}). The nonmagnetic sandwich must give up its $\eta^6$--$\eta^6$ coordination, so the $M=0$ branch rises steeply. The magnetic branch, on which the gap is open, is almost flat from $f=0.2$ to the bridge. The lowest branch changes from $M=0$ to magnetic between $f=\fAAcrossLoThree$ and $\fAAcrossHiThree$. The envelope then follows the magnetic plateau, and its maximum, $E_\mathrm{env}=\EaAAscanThree$~meV, is at \AApeakWhereThree. In this cell the AA barrier is therefore the bridge energy, but the path to it passes through a change of spin state.

\emph{Larger cells.} In a $3\times3$ cell, opening the gap around Cr opens it across the whole cell. In larger cells the opening can stay local, and the bridge energy falls accordingly (Table~\ref{tab:aa}):
\begin{itemize}
\item pinned far field: \dEAAbridgeFourPin~meV ($4\times4$) and \dEAAbridgeFivePin~meV ($5\times5$);
\item free far field: \dEAAbridgeFourFree~meV ($4\times4$) and \dEAAbridgeFiveFree~meV ($5\times5$).
\end{itemize}
The hollow is barely affected: in the $5\times5$ cell the pinned hollow lies only \dEAAhollowPinFreeFiveMeV~meV above the free one.

As the bridge drops, the spin change matters more. In the $4\times4$ pinned cell the lowest branch changes between $f=\fAAcrossLoFour$ and $\fAAcrossHiFour$. Near the change the magnetic branch is a lower-moment substate (total $M\approx4$~$\mu_\mathrm{B}$) that lies above the bridge. A higher-moment substate ($M\approx4.8$~$\mu_\mathrm{B}$) is lower from $f=0.3$ onward. Restarts from it at $f\le0.25$ return to the lower-moment state, and at $f=0.3$--0.5 it lies below the bridge. The envelope peaks at \AApeakWhereFour{} ($f=\fAApeakFour$), at $E_\mathrm{env}=\EaAAscanFour$~meV. The intermediate branch with $M=2$ is never the lowest.

In the $5\times5$ pinned cell the lowest branch changes between $f=\fAAcrossLoFive$ and $\fAAcrossHiFive$, and $E_\mathrm{env}=\EaAAscanFive$~meV at $f=\fAApeakFive$ ($M=\MAApeakFive~\mu_\mathrm{B}$), in \AApeakWhereFive. Beyond the crossing the magnetic branch keeps falling through $f=0.3$, where it lies below the bridge, as in the $4\times4$ cell.

\emph{Free far field.} Relative to $3\times3$, the pinned scans change the cell size and the far-field treatment together. We therefore repeated the scan in the $4\times4$ and $5\times5$ cells with all host heights free (bottom row of Fig.~\ref{fig:aa}). The picture is the same. The lowest branch changes from $M=0$ to magnetic between $f=\fAAcrossLoFourFree$ and $\fAAcrossHiFourFree$ ($4\times4$) and between $\fAAcrossLoFiveFree$ and $\fAAcrossHiFiveFree$ ($5\times5$). The envelope peaks just past the change, at $E_\mathrm{env}=\EaAAscanFourFree$~meV ($f=\fAApeakFourFree$, $4\times4$) and \EaAAscanFiveFree~meV ($f=\fAApeakFiveFree$, $5\times5$). Beyond the change the magnetic branch keeps falling and drops below the relaxed bridge, so on this branch the bridge is not the minimum, consistent with the off-line relaxation noted among the limitations. In the $5\times5$ free cell $E_\mathrm{env}$ lies within a few meV of the bridge energy. This is a coincidence of the sampled values: the peak is at the spin change, not at the bridge. Over the four larger scans, the straight-line crossing estimates differ from $E_\mathrm{env}$ by at most \EaAAinterpMaxDevMeV~meV.

Rows of Table~\ref{tab:aa} with the same far field now isolate the cell size, and rows with the same cell isolate the far field. In the same cell, freeing the far field lowers $E_\mathrm{env}$ by \dEAAfarFourMeV~meV ($4\times4$) and \dEAAfarFiveMeV~meV ($5\times5$). With the same treatment, $E_\mathrm{env}$ falls from $4\times4$ to $5\times5$ in both cases: from \EaAAscanFour{} to \EaAAscanFive~meV pinned and from \EaAAscanFourFree{} to \EaAAscanFiveFree~meV free. The AA barrier is therefore not converged at $5\times5$ in either treatment, and most of the $3\times3$ value (\EaAAscanThree~meV) is a small-cell effect.

\subsection{Sensitivity}
\label{sec:sens}
\begin{table*}
\caption{\label{tab:sens}Numerical and methodological sensitivity of the site energy differences ($3\times3$, meV).
Each variant is a static calculation at the relaxed reference geometry with one setting changed; the entry is the
change of the difference relative to the reference protocol (first column).}
\begin{ruledtabular}
\begin{tabular}{lrrrrr}
Difference & reference & 550~eV & $6\times6$ $k$ & no D3 & 0.05~eV smear \\
\hline
AA bridge$-$hollow & 636 & -2 & 0 & -224 & -- \\
AB topB$-$mid & 103 & +1 & +20 & -41 & -- \\
AB top$-$mid & 158 & -2 & +3 & -118 & -- \\
mono bridge$-$hollow & 50 & +2 & +17 & -132 & -2 \\
AB mid$\to$mid$^\prime$ CI-NEB barrier & 15 & -- & -5 & -22 & -- \\
\end{tabular}
\end{ruledtabular}
\end{table*}

\begin{table*}
\caption{\label{tab:u}Effect of an on-site Hubbard correction on Cr $d$ ($3\times3$). Each state is given relative to the lowest
state found for the same host and treatment (meV), with the total moment $M$ ($\mu_\mathrm{B}$). PBE+U runs are relaxed from the
PBE geometries; ``magnetic seed'' at the AA hollow holds Cr in plane and starts from 4~$\mu_\mathrm{B}$. Empty: not computed.
--: no converged run (at $U=1.5$~eV this state failed self-consistency; in PBE it was not computed separately). All other
entries are converged relaxations. At $U=0.5$ and 1.0~eV only the two AA states that decide the ordering were computed.}
\begin{ruledtabular}
\begin{tabular}{lrrrrrrrrr}
State & PBE & $U=0.5$ & $M$ & $U=1.0$ & $M$ & $U=1.5$ & $M$ & $U=3$ & $M$ \\
\hline
AA hollow, $M=0$ seed & 0 & 0 & 0.0 & 74 & 0.0 & 440 & 0.0 & 1544 & 0.0 \\
AA hollow, magnetic seed & -- &  &  &  &  & -- & -- & 98 & 4.5 \\
AA bridge & 636 & 284 & 4.4 & 0 & 4.5 & 0 & 4.5 & 0 & 4.6 \\
\hline
AB mid & 0 &  &  &  &  &  &  & 0 & 4.5 \\
AB topB & 103 &  &  &  &  &  &  & 22 & 4.6 \\
AB top & 158 &  &  &  &  &  &  & 40 & 4.6 \\
\hline
mono minimum & 0 &  &  &  &  &  &  & 8 & 5.7 \\
mono bridge & 59 &  &  &  &  &  &  & 0 & 5.8 \\
\end{tabular}
\end{ruledtabular}
\end{table*}

\begin{table*}
\caption{\label{tab:size}Cell-size dependence of the site energy differences (meV). ``free'': all host coordinates
relaxed except the registry anchors; ``pin'': carbon heights beyond the pinning radius held at the pristine
interlayer spacing. The two far-field treatments are not guaranteed to bound the dilute limit; their spread is an
uncertainty estimate. The $5\times5$ bridges, the free $4\times4$ AA bridge and the larger-cell surface top sites hold Cr in
plane on the site; the other runs were relaxed freely and stayed within 0.1~\AA{} of it. --: not computed.}
\begin{ruledtabular}
\begin{tabular}{lrrrrr}
Difference & $3\times3$ & $4\times4$ free & $4\times4$ pin & $5\times5$ pin & $5\times5$ free \\
\hline
AA bridge$-$hollow & 636 & 223 & 314 & 245 & 183 \\
AB topB$-$mid & 103 & 123 & -- & -- & -- \\
AB top$-$mid & 158 & 153 & -- & -- & -- \\
mono bridge$-$hollow & 50 & -42 & -- & -- & -39 \\
mono top$-$bridge & 2 & 13 & -- & -- & 10 \\
\end{tabular}
\end{ruledtabular}
\end{table*}

\emph{Numerical settings.} At fixed geometry, raising the cutoff to 550~eV changes the site energy differences by at most a few meV (Table~\ref{tab:sens}). The $6\times6$ $k$-mesh changes them by up to about 20~meV, which matters for the smallest barriers. Table~\ref{tab:sens} therefore also gives the AB CI-NEB barrier itself, recomputed at fixed geometry. With the denser mesh it stays small. Without D3 the midpoint image falls below the endpoint, so in that test the inversion centre is no longer a minimum along the hop.

\emph{Dispersion.} Removing D3 changes every difference by tens to hundreds of meV. For AA the change is \dEAAnoDthreeMeV~meV, so the dispersion cost of opening the AA gallery is a major part of that barrier. On the surface, the bridge falls below the hollow without D3. This is a static test at the PBE+D3 geometries, but it shows that the weakly bound surface state depends on dispersion.

\emph{Hubbard U.} Table~\ref{tab:u} shows that the AA result depends on the treatment of Cr $d$ correlation. The $U=0.5$ and 1.0~eV columns locate the change. The magnetic bridge lies \dEAAbridgeUhalfMeV~meV above the nonmagnetic hollow at $U=0.5$~eV and \dEAAbridgeUoneMeV~meV relative to it at $U=1.0$~eV, so the ordering inverts between these two values. Above that, the AA minimum is no longer the nonmagnetic sandwich. In the AB gallery and on the surface, Cr is magnetic with and without $U$. With $U=3$~eV the AB ordering of sites is kept with smaller differences, and on the surface the bridge becomes the lowest site found in the $3\times3$ cell, as it already is in the larger PBE cells. The AA spin-switch picture, and the AA barriers in Table~\ref{tab:aa}, therefore hold for PBE+D3 and not in general. We do not give AA rates under $U$.

\emph{Cell size.} For AB, the topB$-$mid difference changes little from $3\times3$ to the free $4\times4$ cell (Table~\ref{tab:size}). The surface changes qualitatively. In the $3\times3$ cell the bridge lies \dEmonoBridgeHollowThreeMeV~meV above the centred hollow, and the top only \dEmonoTopBridgeThreeMeV~meV above the bridge. In the $4\times4$ and $5\times5$ cells the bridge lies \dEmonoBridgeHollowFourMeV{} and \dEmonoBridgeHollowFiveMeV~meV relative to the hollow, so it becomes the minimum. In both cells a free relaxation started 0.05~\AA{} off the bridge stays there, within \monoBridgeMinMaxMeV~meV of the bridge energy, and finds no lower point. The top lies \dEmonoTopBridgeFourMeV{} and \dEmonoTopBridgeFiveMeV~meV above the bridge, with the same moment as the bridge ($\MmonoBridgeFive~\mu_\mathrm{B}$ in $5\times5$). To sample the hop, we relaxed points at one and two thirds of the straight line from the bridge to the top, with Cr held in plane. They lie \dEmonoSegAFourMeV{} and \dEmonoSegBFourMeV~meV above the bridge in $4\times4$ and \dEmonoSegAFiveMeV{} and \dEmonoSegBFiveMeV~meV in $5\times5$, so in both cells the energy rises along the line and the top is the highest sampled point. We take this highest sampled point (\dEmonoHopMaxFourMeV{} and \dEmonoHopMaxFiveMeV~meV) as the estimate of the barrier for a bridge$\to$top$\to$bridge hop. Like $E_\mathrm{env}$, it is a sampled value on a straight line, not a converged saddle point. A free relaxation started 0.05~\AA{} off the $5\times5$ hollow leaves it, moving \dxyHollowminFive~\AA{} toward a top site, and stops \dEmonoHollowminFiveMeV~meV above the bridge. The hollow is therefore not a local minimum in this cell, and the search finds no point below the bridge. The $4\times4$ and $5\times5$ values agree to a few meV, so we take the $5\times5$ cell as the surface estimate. The $3\times3$ surface landscape is a small-cell artefact. The likely cause is the Cr--Cr distance of 7.4~\AA, at which the weakly bound adatom still interacts with its periodic images.

\section{Discussion and limitations}

\emph{Rates.} With the barriers above and $\nu=10^{12}$--$10^{13}$~s$^{-1}$ at \KinT~K, the hop rates are:
\begin{itemize}
\item AB gallery: $10^{\logkABlo}$--$10^{\logkABhi}$~s$^{-1}$;
\item surface: $10^{\logkmonolo}$--$10^{\logkmonohi}$~s$^{-1}$ with the $3\times3$ CI-NEB barrier, and $10^{\logkmonoFivelo}$--$10^{\logkmonoFivehi}$~s$^{-1}$ with the highest sampled point of the $5\times5$ hop;
\item AA gallery, if the spin change along the path is not rate-limiting and with the same prefactor: $10^{\logkAAThreelo}$--$10^{\logkAAThreehi}$~s$^{-1}$ with the $3\times3$ barrier, $10^{\logkAAFivelo}$--$10^{\logkAAFivehi}$~s$^{-1}$ with the pinned $5\times5$ estimate, and $10^{\logkAAFiveFreelo}$--$10^{\logkAAFiveFreehi}$~s$^{-1}$ with the free $5\times5$ estimate.
\end{itemize}
In AB stacking at fixed registry, confinement does not slow Cr down: the inversion-centre network is about as mobile as the surface. In AA stacking, Cr is held in the nonmagnetic sandwich longer. At the same prefactor, and with the $3\times3$ AB band as the reference, it hops \dlogkABAAFiveFree--\dlogkABAAFive{} orders of magnitude more slowly than in AB with the $5\times5$ AA estimates, and \dlogkABAAThree{} orders more slowly with the $3\times3$ one. Rates computed in small cells therefore overstate the AA trapping by many orders of magnitude, and the $5\times5$ values are not converged either. The AA hop also requires the change from $M=0$ to a magnetic state. If that change is slow, the AA rate is lower still, so the AA numbers are conditional estimates.

For deposition on graphene--nanotube--graphene junctions~\cite{chen2017ucr}, the results imply that the local stacking registry matters more than confinement as such. Because Cr itself favours AA registry (see Limitations), whether the layers around it keep their Bernal registry decides which of the two cases applies. These rates are order-of-magnitude estimates for an isolated atom and do not model the experiment.

\emph{Comparison with earlier AA work.} Han \emph{et al.} reported 0.74~eV in a $6\times6$ cell~\cite{han2018rscadv}. Our $3\times3$ value (\EaAAscanThree~meV) is close to theirs, but our larger cells give much smaller barriers. Their cell was large, so the difference is not a cell-size artefact on their side. Their calculation differs from ours in the dispersion model (D2), the code and, for the ordered compounds they report, a Cr moment of about 4~$\mu_\mathrm{B}$. Our minimum is nonmagnetic. Given the sensitivity to dispersion and to $U$ found above, we cannot reconcile the two numbers without their inputs. We note only that the spin state of the AA minimum, and how it is followed along the path, can change the barrier by hundreds of meV.

\emph{Limitations.}
\begin{itemize}
\item Spin is treated collinearly without spin--orbit coupling. The AA barrier comes from separately relaxed branches sampled at discrete points along a constrained line. It is an estimate of the collinear barrier along that line. It is not a bound, a minimum-energy crossing point or a nonadiabatic rate.
\item The AA scans hold Cr on the hollow$\to$bridge mirror line. In one unconstrained free $4\times4$ relaxation started at the bridge, the magnetic Cr slid toward the hollow, partly off the line, and reached a lower energy than the bridge before the run was stopped. Paths off the line may therefore lower the AA barrier somewhat.
\item The registry is fixed, and Cr itself favours AA registry. Measured from each pristine bilayer, Cr binds \dEbStackThreeMeV~meV more strongly in AA than in AB in the $3\times3$ cell, \dEbStackFourFreeMeV{} and \dEbStackFiveFreeMeV~meV in the free $4\times4$ and $5\times5$ cells, and \dEbStackFourPinMeV~meV in the pinned $4\times4$ cell. The sign agrees with ordered C$_{12}$Cr, computed with the same code and functional, where Cr binds 0.65~eV more strongly in AA and intercalation reverses the stacking preference by 400--600~meV for the group-6 metals~\cite{chen2026tmbg}. The size is similar in our $3\times3$ cell, which is also a concentrated limit, but in the free cells it levels off near 250~meV. In the $3\times3$ cell the preference outweighs the stacking cost of the whole cell (\dEstackHostThreeMeV~meV). That cost grows with the cell area (\dEstackHostFiveMeV~meV in $5\times5$), and in the free $5\times5$ cell the AA cell with Cr already lies \dEstackCrFiveFreeMeV~meV above the AB one. A single Cr between AB-stacked layers could still lower its energy by pulling only the layers around it into AA registry, if that region and its boundary cost less than the binding gain. Cr would then be held as in the AA gallery, and the AB mobility found here would not apply. Our periodic cells with fixed registry cannot form such a region, so the AB results hold only for Cr that keeps the Bernal registry around it. The preference also depends on $U$: at $U=3$~eV the lowest AA and AB states we sampled give about \dEbStackUthreeMeV~meV, in the $3\times3$ cell, where the PBE value is largest.
\item The dilute limit is approached only up to $5\times5$ cells. Both far-field treatments are used in the $4\times4$ and $5\times5$ cells, and in both the AA barrier still falls between them. The CI-NEB paths are computed only in $3\times3$ cells. For the surface, whose landscape changes beyond $3\times3$, the larger-cell barrier is the highest of a few points sampled on a straight hop, not a band. For AB, the size check covers only the site differences.
\item The sensitivity tests other than cell size are static calculations at fixed geometry. The PBE+U runs are relaxations but do not include paths.
\item Prefactors are not computed. Entry of Cr into the gallery, Cr clustering and defects are outside the model.
\end{itemize}

\section{Conclusions}

With one protocol for all three environments, a single Cr atom is highly mobile on graphene (about \dEmonoHopMaxFiveMeV~meV in a $5\times5$ cell; \EamonoNEB~meV in $3\times3$) and, as long as the layers keep their Bernal registry around it, about as mobile between AB-stacked layers (\EaABNEB~meV). In the AB gallery it moves between the inversion centres of the bilayer, which form a kagome net, while keeping a moment of about \MABmid~$\mu_\mathrm{B}$.

Between AA-stacked layers, PBE+D3 binds Cr in a nonmagnetic sandwich that pulls the layers together. Leaving it requires opening the gallery and switching Cr into a magnetic state, and the spin-resolved barrier along the hollow$\to$bridge line falls from \EaAAscanThree~meV in a $3\times3$ cell to \EaAAscanFive{} and \EaAAscanFiveFree~meV in $5\times5$ cells with pinned and free far field, and is not yet converged. With equal prefactors and if the spin change is not rate-limiting, Cr therefore hops \dlogkABAAFiveFree--\dlogkABAAFive{} orders of magnitude more slowly between AA layers than between AB layers with the $5\times5$ AA estimates and the $3\times3$ AB band, not the \dlogkABAAThree{} that the $3\times3$ AA value would suggest.

Small cells, the dispersion model and the treatment of Cr $d$ correlation each change the AA result substantially. Cr binds \dEbStackFiveFreeMeV--\dEbStackThreeMeV~meV more strongly in AA than in AB registry in our PBE+D3 cells, so whether it pulls AB-stacked layers locally into AA decides which of the two cases applies. Experiments that can tell AA from AB stacking near deposited Cr would test this.

\section*{Data availability}
\label{sec:data}
The ancillary files of this arXiv submission (directory \texttt{DA2\_data}) contain the task definition, trajectories and, for every finished run, the result record of every run used here, the scripts that defined and ran the calculations, and the script that regenerates every number, table and figure in this paper from those records; its \texttt{README.md} gives the commands. The GPAW text logs are available from the corresponding authors on request.

\bibliography{references}

\end{document}